\documentstyle[11pt,newpasp,twoside]{article}
\input{psfig}

\newcommand{\gtae}{\raisebox{-0.6ex}{$\,\stackrel
{\raisebox{-.2ex}{$\textstyle >$}}{\sim}\,$}}
\newcommand{\smallHI}{H{\,\scriptsize I}}
\newcommand{\Halpha}{H$\alpha$}
\newcommand{\HI}{H{\,\small I}}

\def\edcomment#1{\iffalse\marginpar{\raggedright\sl#1\/}\else\relax\fi}
\reversemarginpar

\begin{document}
\title{Large Gas Disks in Radiogalaxies}
 \author{R. Morganti, T.A. Oosterloo, S. Tinti}
\affil{Netherlands Foundation for Research in Astronomy, PO Box 2, 7990
     AA, Dwingeloo, The Netherlands}
\author{C.N. Tadhunter, K.A. Wills}
\affil{Dep. Physics and Astronomy,
             University of Sheffield, S7 3RH, UK}
\author{G. van Moorsel}
\affil{National Radio Astronomy Observatory, Socorro,
             NM 87801, USA}

\begin{abstract}
  
  
  Large-scale (up to 100 kpc) \HI\ structures have been found in three radio
  galaxies using the WSRT and the VLA. In one case, the \HI\ has been detected
  in emission while in the other two \HI\ absorption is detected against the
  radio lobes. In at least two of the three studied radio galaxies the \HI\ 
  appears to be distributed in a large disk and the large amount of neutral
  gas detected ($\gtae 10^9$ M$_{\odot}$) indicates that it is resulting from
  mergers of gas-rich galaxies.  The relatively regular structure and
  kinematics of these disks suggest that the merger must have happened more
  than $10^8$ yrs ago, therefore supporting the idea that the radio activity
  starts late after the merger. In these low redshift radio galaxies we may
  witness the processes that are more efficiently and frequently happening at
  high-$z$.

\end{abstract}

\section{Introduction}

There exists compelling morphological and kinematical evidence that the
activity in powerful radio galaxies is triggered by galaxy mergers and
interactions.  This is also supported by the theoretical results (Kauffmann \&
Haehnelt 2000) that the evolution of supermassive black holes is strongly
linked to the hierarchical build-up of galaxies.  Although these processes are
likely to be more efficient and frequent at high redshifts, they are observed
also  in relatively ``nearby'' radio galaxies.  However,
considerable uncertainties remain about the nature of the triggering events.
Outstanding questions include: 1) is the activity triggered by major mergers
between gas-rich galaxies or by minor accretions events?  2) what is the
relationship with other types of merging systems such as ULIRGs? 3) at what
stage of the merger do the jets and associated activity occur?  4) do all
giant elliptical galaxies go through a radio phase as they evolve via galaxy
interactions?

It is now clear that some early-type galaxies contain a large amount of \HI,
in some cases even 10$^{10}$ M$_{\odot}$ (e.g. Oosterloo et al, 2001 and ref.
therein).  This gas is an important element for understanding the origin and
the evolution of these galaxies. Large tails/arms of neutral hydrogen are a
prototypical signature of a recent merger, while gas settled in large
disk-like structures may indicate an older merger. If radio galaxies have a
similar origin, we may expect to find also there such \HI\ signatures.  The
\HI\ properties could allow us to understand the temporal sequence between
merger, starburst phase and onset of the radio activity.  It is, therefore,
important to make the connection between the presence of a rich ISM, the radio
galaxies and the evolution of the population of giant elliptical galaxies in
general.  The detection of \HI\ around radio galaxies would therefore give a
powerful tool to answer some of the above questions.

\begin{figure}
\centerline{\psfig{figure=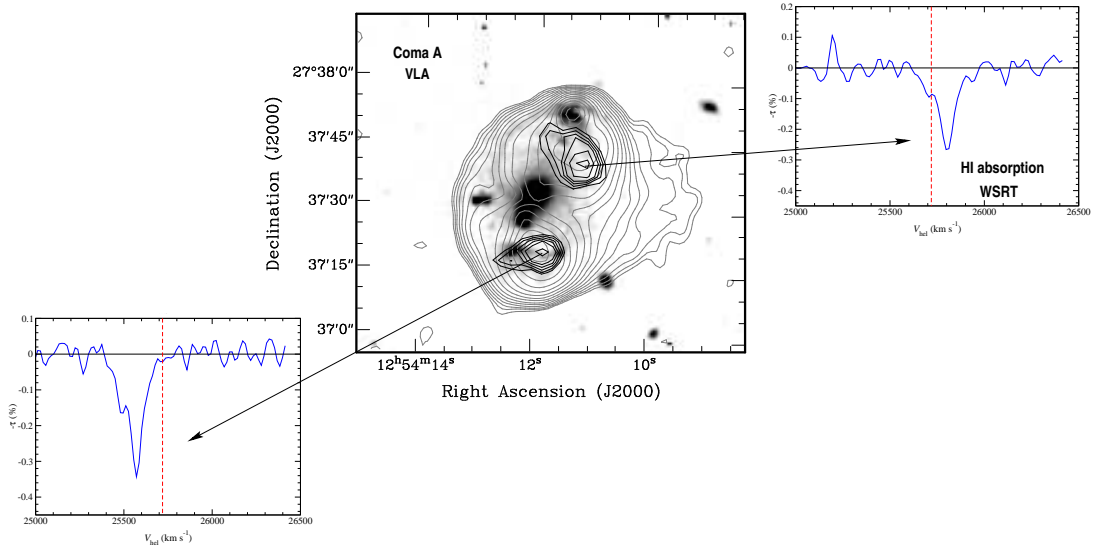,width=15cm,angle=-90}}
\caption{\Halpha\ image (grey scale) of Coma~A (from Tadhunter et al.\ 2000)
 with superimposed contours of a 20-cm continuum image (about
 1$^{\prime\prime}$ resolution, from van Breugel et al. 1985). The dark contours
 represent  the region where \HI\ absorption has been detected. The profiles of the optical depth ($\tau$) of the
\smallHI\ absorption against the two radio lobes.}
\end{figure}

\section{Radiogalaxies with extended \HI\ disks}

So far we have found three radio galaxies where the neutral hydrogen is
detected at large distances from the nucleus. An other possible candidate is
the radio galaxy 3C~234 studied by Pihlstr\"om (2001).

\subsection{Coma~A}

Coma~A ($z = 0.08579$) has a spectacular system of interlocking arcs and filaments detected in
optical emission lines (see Fig.\ 1 and Tadhunter et al.  2000 for details).
This ionised gas and the radio structure show a striking match that is
suggestive of a complex interaction between the radio structure and a rich
interstellar medium.  Using the WSRT and the VLA we have detected \HI\ 
absorption in front of both radio lobes of Coma A (see Fig.\ 1 and Morganti et
al. 2001).  This is a rare case where the absorption is not detected against
the nuclear regions of a radio galaxy, but it is situated at large distances
(30 kpc) from the centre.  The kinematics of the neutral and ionised gas
suggests that they are part of the same structure, likely a large-scale disk
with the radio lobes expanding into this disk.

\begin{figure}
\centerline{\psfig{figure=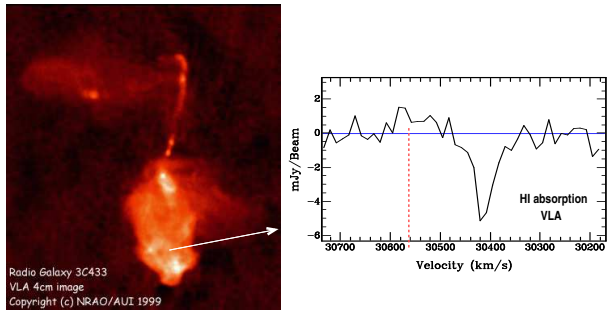,width=8cm,angle=-90}}
\caption{({\sl Left}) Continuum image of 3C~433 (from Black et al. 1992) and \HI\
  absorption profile ({\sl Right}). The systemic velocity is indicated.}
\end{figure}

\subsection{3C~433}

3C~433 is a radio galaxy ($z = 0.1016$) with an unusual double-lobed radio
morphology (see Fig. 3).  We have observed 3C~433 using the VLA (C-array) and
we have found that at least part of the \HI\ absorption (originally detected
with the Arecibo telescope, Mirabel 1989) is situated against the southern
radio lobe at about 40 kpc from the nucleus (see Fig.\ 2). The optical depth
of the absorption is only about 2\%.  Unlike Coma A, no ionised gas has been
detected near the location of the \HI\ absorption.
3C~433 is a far-IR bright radio galaxy with a young stellar population
component (Wills et al.  2001).  The presence of large-scale neutral hydrogen
could be related to the particular stage of the evolution of this radio
galaxy.  3C~433 is perhaps a relatively young radio galaxy.

\subsection{B2~0648+27}

B2~0648+27 ($z = 0.041$) is a compact radio galaxy and in this object we have
detected neutral gas both in emission and in absorption (see Fig.\ 3, WSRT
observations).  The neutral gas is in a disk-like structure of about 100 kpc
in size and contains about $10^{10}$ M$_{\odot}$ of \HI.  Such a large amount
of gas is usually believed to originate from a "major merger", i.e.  a merger
of two large disk galaxies.  Like 3C~433, B2 0648+27 is a far-IR bright galaxy
with a young stellar population component.  The relatively regular kinematics
of the gas indicates that the merger must have happened more than $10^8$ yrs
ago and therefore the radio activity (usually estimated to last for few time
$10^7$ yrs) appeared at a late stage in the merger. \HI\ in absorption is
detected only in the centre, where the radio continuum is present, and has a
kinematics similar to the molecular gas detected, in large amounts, in this
object (Mazzarella 1996).

\begin{figure}
\centerline{\psfig{figure=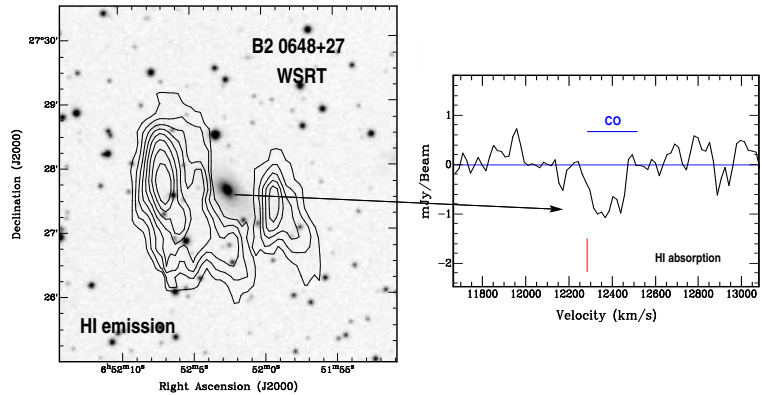,width=10cm,angle=-90}}
\caption{({\sl Left}) \HI\ total intensity contour  of
  the radio galaxy B2 0648+27 superimposed on to an optical image. ({\sl
    Right}) \HI\ absorption profile (the optical systemic
  velocity is marked).  The range of the CO emission (from Mazzarella et al.
  1993) is also indicated.}
\end{figure}

\section{Summary}

Large-scale (up to 100 kpc) gas structures have been found in three radio
galaxies and they can  be used to understand the origin and the evolution of these
systems. The presence of a young stellar population component in
two of these objects will give further constrains on type and age of the
merger.  Extended \HI\ absorption 
(observed against the Ly$\alpha$ emission) has been found in a high fraction
of high-$z$ radio galaxies (van Ojik et al.  1997). This is considered an
indication that high-$z$ radio galaxies are located in dense environments and helps in probing
the effects of radio jet propagation in this medium. Although this may be happening more
efficiently and frequently at high redshifts, in the low redshift
radio galaxies described here we may witness a similar situation. However, in
the low redshift galaxies, we will be able to investigate
in much more  detail the relation between the thermal and non-thermal gas.

\end{document}